\documentclass[reprint,jap,aip]{revtex4-1}
\usepackage{amssymb}
\usepackage{amsmath}
\usepackage{graphicx}
\usepackage{dcolumn}
\usepackage{bm}
\usepackage[latin1]{inputenc}

\newcommand{\bra}{\langle}
\newcommand{\ket}{\rangle}
\begin{document}

\title[Magnetoresistance and transistor-like behavior of a DQD via Andreev
reflections]{Magnetoresistance and transistor-like behavior of a double
quantum-dot via crossed Andreev reflections}
\author{E. C. Siqueira}
\email{ecosta@dfq.feis.unesp.br}
\affiliation{Departamento de Física e Química,\\Universidade Estadual Paulista --- UNESP,\\
Av. Brasil, Centro, 15385-000 - Ilha Solteira, SP - Brasil}
\author{G. G. Cabrera}
\email{cabrera@ifi.unicamp.br}
\affiliation{Instituto de Física  `Gleb Wataghin',
Universidade Estadual de Campinas (UNICAMP),\\
Campinas 13083-859, SP, Brasil}
\date{\today}

\begin{abstract}
The electric current and the magnetoresistance effect are studied in a
double quantum-dot system, where one of the dots $QD_{a}$ is coupled to two
ferromagnetic electrodes $(F_{1},F_{2})$, while the second $QD_{b}$ is
connected to a superconductor $S$. For energy scales within the
superconductor gap, electric conduction is allowed by Andreev reflection
processes. Due to the presence of two ferromagnetic leads, non-local crossed
Andreev reflections are possible. We found that the magnetoresistance sign
can be changed by tuning the external potential applied to the ferromagnets.
In addition, it is possible to control the current of the first ferromagnet (%
$F_{1}$) through the potential applied to the second one ($F_{2}$). We
have also included intradot interaction and gate voltages at each quantum dot and
analyzed their influence through a mean field approximation. The interaction
reduces the current amplitudes with respect to the non-interacting case,
but the switching effect still remains as a manifestation of quantum coherence,
in scales of the order of the superconductor coherence length.
\end{abstract}

\maketitle





\section{Introduction}

The study of transport properties of hybrid nanostructures is a very active field of research, involving new and interesting physical phenomena that appear at the nanometer scale, with great potential for developing future technology in mesoscopic systems. Within this context, systems based on combinations of
superconductors and ferromagnets are particularly interesting, since the
interplay between these two phenomena can give rise to unusual effects. It
is well known that in ferromagnetic/superconducting ($F/S$) junctions the
conductance can be controlled through the ferromagnet polarization. For
energies within the superconductor gap, the conduction process is
established via Andreev reflections \cite{andreevref} (AR). In this process,
two electrons of $F$ with opposite spins recombine into a Cooper pair in $S$
(with total spin $S=0$). The Cooper pairs are the supercurrent carriers of
the superconductor and these pairs are highly correlated in large distances
in comparison to the interatomic distances. This feature has been explored
by Deutscher and Feinberg \cite{deutscher} to propose a non-local Andreev
reflection (called crossed AR), where two electrons of different leads can
combine into a Cooper pair if the distance between these leads is smaller
than the coherence length. Since this proposal, there has been a profusion
of works exploring crossed AR in different geometries \cite%
{Circuit,Benjamin,Peysson,Giazotto,DeutscherCross,Melin,Citro,Bai1,Bai2},
resonant nanostructures involving quantum dots \cite%
{spininjector,Xiufeng,Feng,Sun,Dolcini,Songintra}, different conduction
regimes \cite{Brinkman,Lambert4} (ballistic and diffusive), and addressing
more fundamental questions, e.g., the entanglement of the quasiparticles in
different leads \cite{Lesovik,feinberg1,feinberg2,feinberg3,Bignon,Yang}.
Within this vast set of hybrid nanostructures, systems composed by double
quantum dots (QDs) are very promising, since this association can serve as a
model of diatomic molecules \cite{Wiel,liu:referee,giavaras:referee}. Many
works involving double QDs have been developed mainly concerning the Kondo
effect \cite{Guo,Eto,Aguado}, scattering with spin inversion \cite{Guo2},
effects of different geometries \cite{suk,Choi} (series and parallel
association), spin detectors \cite{Giavaras:2,Chorley} and systems involving
superconductors. In the latter case, there are studies involving Josephson
molecular junctions \cite{zhang,Buitelaar,Buitelaar2} and transport by AR
\cite{Kawakami,Tanaka}.

\begin{figure}[h]
\centering
\includegraphics[scale=0.60]{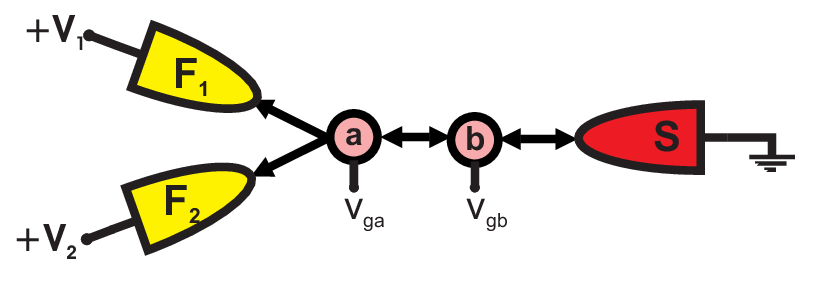}
\caption{(Color Online)~Schematic
diagram showing the
(\textit{F}$_{1}$,\textit{F}$_{2}$)-\textit{QD}$_{a}$-\textit{QD}$_{b}$-\textit{S} system. The magnetization of $F_{1}$ is assumed to be fixed and
the magnetization of $F_{2}$ can be varied for an angle $\theta$ with
respect to the $F_{1}$ magnetization. $V_{1}$ and $V_{2}$ are the external
potentials applied to  $F_{1}$ and $F_{2}$, respectively while the
superconductor is grounded. Gate voltages are also applied to the QDs, with
$V_{ga}$ and $V_{gb}$ being the potentials applied to $a$ and  $b$, respectively.}
\label{esquema}
\end{figure}

By considering the outstanding properties of crossed ARs and the promising
feature of double QDs, we propose a prototype of a molecular transistor by
combining two QDs with a superconductor and two ferromagnetic electrodes. A
schematic diagram of the system is shown in Fig. \ref{esquema}. There are
two ferromagnetic electrodes, $F_{1}$ and $F_{2}$, attached to the first QD
and a superconductor electrode is connected to the second one. The dot
coupled to the ferromagnetic electrodes ($F$) is called $a$, and $b$ is the
one coupled to the superconductor ($S$). The superconductor has its chemical
potential fixed to zero, and independent voltage bias are applied to the
ferromagnets which are called $V_{1}$ and $V_{2}$. There are also gate potentials applied to the dots, denoted by $V_{ga}$ and $V_{gb}$. By exploring the resonant
structure of the local density of states (LDOS) and the non-local feature of
the crossed AR, we show that is possible to switch the current at one
ferromagnetic lead by the applied bias in the second one. In addition, the
magnetoresistance sign can also be changed through the bias. The control of
the current via external parameters can be of interest in applications of
molecular electronics.

We assume the existence of an intradot interaction at each QD and use a
mean-field approximation to include its effect in our calculation. However,
we have not considered the occurrence of Kondo resonances at the QD's. While
the Kondo effect has been experimentally observed in semiconducting QDs,
coupling the dot to a ferromagnetic electrode will split the dot level,
leading to the suppression of the Kondo effect\cite{martinek1,martinek2}.
Electron pairing in the superconductor electrode also competes with Kondo
through the proximity effect\cite{bergeret}. Now, a discussion about the
relative magnitude of the correlation parameters is in order. In this paper,
$\mathcal{U}$ is limited to the gap value, since we analyzed the
contribution of a pure Andreev current (subgap current). Our study is then
confined to the weak correlation regime. This also restricts the voltages to
very small values, typically of the order of mV or smaller. In fact, in the
experiment by Beckmann \cite{Beckmann} \textit{et. al.}, the superconductor
gap of the $Al$ film with thickness of 80~nm was found to be $\sim0.18$~meV.
In another experiment performed by Russo \cite{Russo} \textit{et. al.},
using $Nb$ films, the superconductor gap obtained was in the rage of 0.90 and
1.45~meV, for films with thicknesses between 15 and 50~nm, respectively.

\section{Model and Formulas}

The system displayed in Fig. \ref{esquema} is described by the following
Hamiltonian:
\begin{equation*}
\mathcal{H}=\mathcal{H}_{1}+\mathcal{H}_{2}+\mathcal{H}_{S}+\mathcal{H}%
_{dqd}+\mathcal{H}_{T}
\end{equation*}%
where the ferromagnet $F_{1}$ is modeled by the Stoner model \cite{fazekas}
given by
\begin{equation*}
\mathcal{H}_{1}=\sum_{k\sigma }\epsilon _{1k\sigma }\hat{a}_{k\sigma }^{\dag
}\hat{a}_{k\sigma }
\end{equation*}%
with $\epsilon _{1k\sigma }=\epsilon _{k}-\text{sgn}(\sigma )h_{1}-\mu
_{1}$. In the same way, the lead $F_{2}$ is described by
\begin{equation*}
\mathcal{H}_{2}=\sum_{k\sigma }\epsilon _{2k\sigma }\hat{b}_{k\sigma }^{\dag
}\hat{b}_{k\sigma }-\sum_{k\sigma }h_{2}\sin \theta \hat{b}_{k\sigma }^{\dag
}\hat{b}_{k\bar{\sigma}}\ ,
\end{equation*}%
with $\epsilon _{2k\sigma }=\epsilon _{k}-\text{sgn}(\sigma )h_{2}\cos
\theta -\mu _{2}$.

The spin bands of $F_{1} (F_{2})$ are split by the exchange energy $%
h_{1}(h_{2})$ and the magnetization direction of $F_{2}$ has an angle $%
\theta $ with respect to the magnetization of $F_{1}$. By changing the value
of $\theta$, we can change the configuration of the system from parallel
alignment ($\theta=0$) to an antiparallel alignment ($\theta=\pi$).

The superconductor is described by the BCS Hamiltonian\cite{BCS},
\begin{equation*}
\mathcal{H}_{S}=\sum_{k\sigma }\epsilon _{k}s_{k\sigma }^{\dag }\hat{s}%
_{k\sigma }+\sum_{k}[\Delta \hat{s}_{k\uparrow }^{\dag }\hat{s}%
_{-k\downarrow }^{\dag }+\text{H.c.}]\ ,
\end{equation*}%
with $\Delta $ being the superconductor gap and the operator $\hat{s}%
_{k\uparrow }^{\dag }\hat{s}_{-k\downarrow }^{\dag }$ creates a Cooper pair
in $S$ . Therefore, we are considering here a conventional singlet
superconductor with $s$-wave pairing symmetry.

The chemical potentials $\mu_{1}$ and $\mu_{2}$ of $F_{1}$ and $F_{2}$ are
fixed by the applied bias $V_{1}$ and $V_{2}$ while the superconductor
chemical potential ($\mu_{S}$) is set to zero as the ground.

The QDs are modeled by the mean-field Hamiltonian,
\begin{align}  \label{model:eq5}
\mathcal{H}_{dqd}=\sum_{\sigma}E_{a\sigma}\hat{n}_{a\sigma}+\sum_{%
\sigma}E_{b\sigma}\hat{n}_{b\sigma}
\end{align}
where $E_{a\sigma}=E_{a}-eV_{ga}+\mathcal{U}\langle\hat{n}_{a\bar{\sigma}%
}\rangle/2$ and $E_{b\sigma}=E_{b}-eV_{gb}+\mathcal{U}\langle\hat{n}_{b\bar{%
\sigma}}\rangle/2$. The QDs energy levels, $E_{a\sigma}$ and $E_{b\sigma}$,
are renormalized by the intradot interaction $\mathcal{U}$. This interaction
couples the energy levels to the mean occupations $\langle\hat{n}_{a\bar{%
\sigma}}\rangle$ and $\langle\hat{n}_{b\bar{\sigma}}\rangle$. In addition,
gate voltages $V_{ga}$ and $V_{gb}$ allow one to tune the position of the
bare QD levels with respect to superconductor chemical potential.

The tunneling between the QDs and leads is described by
\begin{multline}
\mathcal{H}_{T}=\sum_{k\sigma }[t_{1}\hat{a}_{k\sigma }^{\dag }\hat{c}%
_{a\sigma }+\text{H.c.}]+\sum_{k\sigma }[t_{2}\hat{b}_{k\sigma }^{\dag }%
\hat{c}_{a\sigma }+\text{H.c.}]  \label{model:eq6} \\
+\sum_{k\sigma }[t_{s}\hat{s}_{k\sigma }^{\dag }\hat{c}_{b\sigma }+\text{%
H.c.}]+\sum_{\sigma }[t_{ab}\hat{c}_{a\sigma }^{\dag }\hat{c}_{b\sigma }+%
\text{H.c.}]\ ,
\end{multline}%
where the last term accounts for the hopping between the QD's. For
simplicity, we have assumed that the hopping matrix elements are independent
of the spin index. This will safely cover the case of homogeneous \
`monodomain' ferromagnets, with no spin-flip scattering. Study of more
general situations for the $S/F$ interface shows the possibility of inducing
a \ `triplet proximity effect' in the ferromagnet from scattering by
inhomogeneities at the interface or at domain walls in the ferromagnet\cite%
{review,referee4_0,referee4_1,referee4_2,referee4_3,referee4_4,referee4_5}. In this
latter case, anomalous Andreev reflections, \emph{i.e.} reflections with
spin-flip into the triplet state, have to be considered as also contributing
to the current. In the present study, the above phenomenon will not be taken
into account, restricting our calculation to ideal homogeneous leads. Note
that in real experimental setups, small magnetic fields can remove any
domain structure.

In order to calculate the transport properties we have used the
non-equilibrium Green's function method \cite{rammer}. All the physical
quantities can be cast in terms of the Green's function of the QD's. Since
we are dealing with ferromagnet and superconductor order parameters, it is
convenient to introduce the Nambu representation by using a generalized
four-dimensional spinor $\hat{\boldsymbol{\Psi }}_{i}=(\hat{c}_{i\uparrow
}^{\dagger }\quad \hat{c}_{i\downarrow }\quad \hat{c}_{i\downarrow
}^{\dagger }\quad \hat{c}_{i\uparrow })^{\dagger }$ with $i=a,b$. This
allows one to treat both order parameters on the same footing.

In terms of Nambu spinors the lesser ($\mathbf{G}^{<}$) and
retarded/advanced Green function $\mathbf{G}^{r/a}$ of QDs are written as
\begin{align}  \label{model:eq6a}
\mathbf{G}^{<}_{ii}(t_{1},t_{2})=i\langle\hat{\boldsymbol{\Psi}}%
_{i}(t_{1})\otimes\hat{\boldsymbol{\Psi}}^{\dagger}_{i}(t_{2})\rangle
\end{align}
and
\begin{multline}  \label{model:eq6b}
\mathbf{G}^{r/a}_{ii}(t_{1},t_{2})=\mp i\vartheta(\pm t_{1}\mp t_{2})\langle%
\hat{\boldsymbol{\Psi}}_{i}(t_{1})\otimes\hat{\boldsymbol{\Psi}}%
^{\dagger}_{i}(t_{2}) \\
+\hat{\boldsymbol{\Psi}}^{\dagger}_{i}(t_{2})\otimes\hat{\boldsymbol{\Psi}}%
_{i}(t_{1})\rangle
\end{multline}
with $i=a,b$. Similiar definitions are given for the leads Green functions.
However, all the physical quantities are determined from the Green functions
of the QDs.

Under stationary regime, the current through the system is time-independent
and we can work with the Fourier transform of the Green functions. In this
case, the total electrical current coming from the ferromagnets and injected
into the superconductor is given by
\begin{align}  \label{model:eq7}
I=\frac{e}{h}\int d\omega[ \mathbf{G}^{r}_{aa}(\omega)\mathbf{\Sigma}%
_{F}^{<}(\omega)+ \mathbf{G}^{<}_{aa}(\omega)\mathbf{\Sigma}_{F}^{a}(\omega)+%
\text{H.c.}]_{11+33}.
\end{align}

The subscript \textquotedblleft 11+33" means taking the sum of 11 and 33
elements of the $4\times 4$ matrix. By adopting the equation of motion
method, the Green function of the dot $a$ has been determined:
\begin{equation*}
\mathbf{G}_{aa}^{r}=\mathbf{G}_{aa}^{r0}+\mathbf{G}_{aa}^{r}\mathbf{t}%
_{ab}^{\dag }\mathbf{G}_{bb}^{r0}\mathbf{t}_{ab}\mathbf{G}_{aa}^{r0}\ ,
\end{equation*}%
with $\mathbf{G}_{aa}^{r0}=\mathbf{g}_{aa}^{r}(\mathbf{1}-\mathbf{\Sigma }%
_{F}^{r}\mathbf{g}_{aa}^{r})^{-1}$~and~$\mathbf{G}_{bb}^{r0}=\mathbf{g}%
_{bb}^{r}(\mathbf{1}-\mathbf{\Sigma }_{S}^{r}\mathbf{g}_{bb}^{r})^{-1}$.

In these equations $\mathbf{G}^{r}_{aa}$ is the Green's function of the
quantum dot $a$; $\mathbf{G}^{r}_{bb}$ is the Green's function of the
quantum dot $b$; $\mathbf{g}^{r}_{aa}$ and $\mathbf{g}^{r}_{bb}$ are the
Green's functions of the dots $a$ and $b$ isolated from the electrodes; $%
\mathbf{t}_{ab}$ describes the coupling between the dots; $\mathbf{\Sigma}%
^{r}_{F}=\mathbf{\Sigma}^{r}_{1}+\mathbf{\Sigma}^{r}_{2}$ and $\mathbf{\Sigma%
}^{r}_{S}$ are the retarded self-energies describing the coupling of the
dots with the ferromagnetic and superconductor electrodes, respectively.
Explicitly, these self-energies are written as,
\begin{equation}  \label{sigmaF}
\mathbf{\Sigma}_{F}^{r,a}(\omega)=\mp\frac{i}{2} \left[%
\begin{array}{cccc}
A_{\uparrow} & 0 & B & 0 \\
0 & A_{\downarrow} & 0 & B \\
B & 0 & A_{\downarrow} & 0 \\
0 & B & 0 & A_{\uparrow}%
\end{array}%
\right],
\end{equation}
with $A_{\sigma}\equiv\Gamma_{1\sigma}+c^{2}\Gamma_{2\sigma}+s^{2}\Gamma_{2%
\bar{\sigma}}$, $B=sc(\Gamma_{2\uparrow}-\Gamma_{2\downarrow})$, $%
s\equiv\sin\theta/2$ and $c\equiv\cos\theta/2$. We also have defined $%
\Gamma_{i\sigma}=2\pi|t_{i}|^{2}N_{i\sigma}$, (with $i=1,2$) as the coupling
strength, with $t_{i}$ being the tunneling amplitude and $N_{i\sigma}$ the
density of states for the ferromagnet spin $\sigma$ band.

The retarded/advanced self-energy of the superconductor is given by,
\begin{equation}  \label{selfenergy:superconductor}
\mathbf{\Sigma}^{r,a}_{S}(\omega)=\mp\frac{i}{2}\Gamma_{s}\rho(\omega) \left[%
\begin{array}{cccc}
1 & -\Delta/\omega & 0 & 0 \\
-\Delta/\omega & 1 & 0 & 0 \\
0 & 0 & 1 & \Delta/\omega \\
0 & 0 & \Delta/\omega & 1%
\end{array}%
\right],
\end{equation}
where $\Gamma_{s}=2\pi |t_{s}|^{2}N_{s}$, with $N_{s}$ being the density of
states of the superconductor in the normal state and $\rho$ is the modified
BCS density of states
\begin{align}  \label{model:eq8a}
\rho(\omega)\equiv\frac{|\omega|\vartheta(|\omega|-\Delta)}{\sqrt{%
\omega^{2}-\Delta^{2}}} +\frac{\omega\vartheta(\Delta-|\omega|)}{i\sqrt{%
\Delta^{2}-\omega^{2}}}
\end{align}
with the imaginary part accounting for Andreev states within the gap \cite%
{claro}.

It is important to note that the definition of $\hat{\boldsymbol{\Psi }}_{i}$
is the same as the one used in Refs. \onlinecite{principal,claro,claro2}. As
a result, the self-energies given by Eqs. \eqref{sigmaF} and
\eqref{selfenergy:superconductor} are the same as those found in Refs. \onlinecite{principal,claro,claro2}.

The ``lesser" Green's function is obtained through the Keldysh equation
\begin{align}  \label{model:eq9}
\mathbf{G}^{<}_{aa}(\omega)= \mathbf{G}^{r}_{aa}(\omega)\mathbf{\Sigma}%
_{Ta}^{<}(\omega)\mathbf{G}^{a}_{aa}(\omega)
\end{align}
with $\mathbf{\Sigma}_{Ta}^{<}(\omega)=\mathbf{\Sigma}_{F}^{<}(\omega)+
\mathbf{t}^{\dag}_{ab}\mathbf{G}^{r0}_{bb}\mathbf{\Sigma}_{S}^{<}(\omega)%
\mathbf{G}_{bb}^{a0}(\omega) \mathbf{t}_{ab}$.

The self-energies $\mathbf{\Sigma}_{F}^{<}=\mathbf{\Sigma}_{1}^{<}+\mathbf{%
\Sigma}_{2}^{<}$ and $\mathbf{\Sigma}_{S}^{<}$ are obtained by the
fluctuation-dissipation theorem $\mathbf{\Sigma}_{i}^{<}=\mathbf{F}%
_{i}(\omega)[\mathbf{\Sigma}_{i}^{a}-\mathbf{\Sigma}_{i}^{r}]$, where $i=1,2$
or $s$. The Fermi matrix $\mathbf{F}_{i}$ is given by,
\begin{equation}  \label{sigmas}
\mathbf{F}_{i}(\omega)=\left[
\begin{array}{cccc}
f_{i} & 0 & 0 & 0 \\
0 & \bar{f_{i}} & 0 & 0 \\
0 & 0 & f_{i} & 0 \\
0 & 0 & 0 & \bar{f_{i}} \\
\end{array}%
\right]
\end{equation}
in which the Fermi functions are defined as $f_{i}=f(\omega-eV_{i})$ and $%
\bar{f_{i}}=f(\omega+eV_{i})$ for $i=1,2$ and $f_{i}=f(\omega)$, if $i=s$.

Since the Green's functions are dependent on mean occupations through the
intradot interaction, it is necessary to calculate those quantities at the
dots. From the definition of the \textquotedblleft lesser" Green's function,
one straightforwardly obtains the system of equations below:
\begin{align*}
\left\bra n_{a\uparrow}\right\ket=\dfrac{1}{2\pi i}\int_{-\infty}^{+%
\infty}G^{<}_{aa,11}[\omega,\left\bra n_{a\uparrow}\right\ket,\left\bra
n_{a\downarrow}\right\ket,\left\bra n_{b\uparrow}\right\ket,\left\bra
n_{b\downarrow}\right\ket] \\
\left\bra n_{a\downarrow}\right\ket=\dfrac{1}{2\pi i}\int_{-\infty}^{+%
\infty}G^{<}_{aa,33}[\omega,\left\bra n_{a\uparrow}\right\ket,\left\bra
n_{a\downarrow}\right\ket,\left\bra n_{b\uparrow}\right\ket,\left\bra
n_{b\downarrow}\right\ket]
\end{align*}
\begin{align*}
\left\bra n_{b\uparrow}\right\ket=\dfrac{1}{2\pi i}\int_{-\infty}^{+%
\infty}G^{<}_{bb,11}[\omega,\left\bra n_{a\uparrow}\right\ket,\left\bra
n_{a\downarrow}\right\ket,\left\bra n_{b\uparrow}\right\ket,\left\bra
n_{b\downarrow}\right\ket] \\
\left\bra n_{b\downarrow}\right\ket=\dfrac{1}{2\pi i}\int_{-\infty}^{+%
\infty}G^{<}_{bb,33}[\omega,\left\bra n_{a\uparrow}\right\ket,\left\bra
n_{a\downarrow}\right\ket,\left\bra n_{b\uparrow}\right\ket,\left\bra
n_{b\downarrow}\right\ket]
\end{align*}

These integral equations must be solved numerically in a self-consistent
way. Once the occupation numbers are obtained, it is possible to calculate
the other physical quantities.

By using the relations above, it is possible to determine the electrical
current as a function of the applied bias $V_{1}$ and $V_{2}$, the magnetization
angle $\theta$ and the gate potentials, $V_{ga}$ and $V_{gb}$. The total
current is obtained by adding the two currents from both ferromagnets. These
currents are summed in the QDs and injected into the superconductor by means of
the Andreev reflection. In this process, an incident electron coming from the
ferromagnetic lead, with energy $\omega$ and spin $\sigma$, combines with a
second electron with energy $-\omega$ and spin $-\sigma$. Both electrons
enter the superconductor as a Cooper pair, leaving a reflecting hole with
spin $-\sigma$ in the ferromagnetic electrode. Since we need both spins
to create a Cooper pair, the AR is prohibited when the
polarization of the ferromagnetic lead is equal to unity. In the setup under
consideration (see Fig. \ref{esquema}), the Andreev reflection may be local,
that is occurring in the same lead of the incident electron, or may be nonlocal,
with the reflected hole appearing in the other lead. For instance, an
incident electron in $F_{1}$ can be reflected as a hole at $F_{1}$ or $F_{2}$.
The latter case, called crossed AR, is possible only if the distance
between $F_{1}$ and $F_{2}$ is of the order of or less than the superconductor
coherence length. Recent experiments probing crossed ARs, estimate superconducting
coherence lengths in the range of 10-15~nm for Nb films\cite{Russo}, and 200-300~nm
for Al films\cite{Beckmann}, depending on the sample, but showing that the effect can
be checked experimentally within the present state of the art in nanodevices.
Crossed ARs allow us to control the current through the angle $\theta$ and the
polarization of the ferromagnets. The polarization is defined in terms of
the coupling constants:
\begin{equation*}
P_{i}=\frac{\Gamma_{i\uparrow}-\Gamma_{i\downarrow}}{\Gamma_{i\uparrow}+%
\Gamma_{i\downarrow}},\qquad i=1,2.
\end{equation*}
The most interesting case is the one when both leads are full polarized.
In this case, AR in the same electrode is not possible and the crossed AR is
the only mechanism to carry current through the system\cite{noanomalous}. As a result, the
current can be tuned from zero to its maximum by varying the angle $\theta$
of the magnetization of $F_{2}$. In fact, when $\theta=0$, the total current
is zero since we have the same spin in both electrodes which implies no
availability of states for the reflected hole. On the other hand, when $%
\theta=\pi$, all the electrons of $F_{1}$ are up-spin and the electrons of $%
F_{2}$ are down-spin and the current exhibits a maximum value. Materials with
a high degree of spin polarization are currently being used to study spin-dependent
transport properties. The most promising case corresponds to $CrO_{2}$, which has
been predicted to be half-metallic and $100\%$ polarized at the Fermi level\cite{halfmetallic}.

In order to compare the current in these two different configurations, we
define the Andreev magnetoresistance as:
\begin{equation}  \label{armr:definition}
ARMR=\frac{|I_{AP}|-|I_{P}|}{|I_{AP}|+|I_{P}|}
\end{equation}
in which $I_{AP}=I(\theta=\pi)$ and $I_{P}=I(\theta=0)$.

The definition (\ref{armr:definition}) is different from the usual one, since we
use the absolute value of the currents. This definition allows us to compare
the amplitude of the currents in terms of the bias of each electrode. In
this system the sign of current in each ferromagnetic terminal is linked to
the averaged chemical potential of the two leads. Thus, it contains the case
that $V_{1}>0$ and $V_{2}<0$ but $I>0$. This unusual behavior is
characteristic of the crossed AR reflection and has been first pointed out
by Y. Zhu \textit{et al.} in a one-dot three-terminal system \cite{principal}%
. By using the definition (\ref{armr:definition}) we can determine which
current is larger through the sign of $ARMR$, even in cases when we
consider the dependence of $ARMR$ with the bias $V_{1}$ or $V_{2}$.

\section{Results and Discussion}

\begin{figure*}[th]
\centering
\includegraphics[scale=0.75]{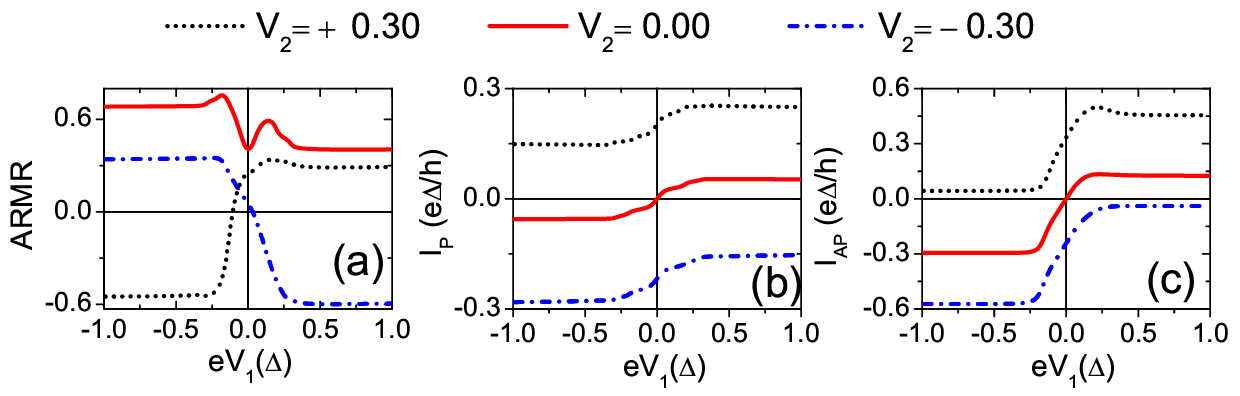}
\caption{(Color Online)~Magnetoresistance $ARMR$ and corresponding currents $%
I_{P}$ and $I_{AP}$ through the system for different values of the applied
bias $V_{1}$ and $V_{2}$. (a) $ARMR$. (b) $I_{P}=I(\protect\theta=0)$. (c) $%
I_{AP}=I(\protect\theta=\protect\pi)$. Fixed parameters: $\Gamma_{1}=0.20$, $%
\Gamma_{2}=0.80$, $\Gamma_{s}=0.30$, $t_{ab}=0.20$, $V_{ga}=V_{gb}=0$, $%
\mathcal{U}=0.40$, $P_{1}=P_{2}=0.95$, $k_{B}T=0.01$. All the parameters are
expressed in superconductor gap units.}
\label{cplpartI}
\end{figure*}

Some $ARMR$ curves are presented in Fig. \ref{cplpartI}a for different
values of the applied bias in the electrodes $F_{1}$ and $F_{2}$. For $%
V_{2}=0$, $ARMR$ is positive in the entire range of $V_{1}$ with a rapid
oscillation around $V_{1}=0$. For $V_{2}=+0.30$ the $ARMR$ displays a
step-like behavior with positive values for $V_{1}>0$ and negative values
for $V_{1}<0$. The trend is inverted for $V_{2}=-0.30$. These results
indicate that one can control the sign of the system magnetoresistance
through external parameters $V_{1}$ and $V_{2}$. In order to understand the $%
ARMR$ curves, in Figs. \ref{cplpartI}b and \ref{cplpartI}c the corresponding
$I_{P}$ and $I_{AP}$ curves are shown. In the parallel configuration, the
total current $I_{P}$ is very small since the polarization values ($%
P_{1}=P_{2}=0.95$) are close to unity. In this case, the crossed AR does not
contribute significantly since the magnetizations of $F_{1}$ and $F_{2}$ are
pointing in the same direction. When the magnetization of $F_{2}$ is
inverted, the crossed AR dominates the conduction process making $I_{AP}$
much higher than $I_{P}$. In this way, when the polarization is close to
unity, the usual situation is to find positive values of $ARMR$ (see Eq. %
\eqref{armr:definition}) since the current $I_{AP}$ is mainly carried by the
crossed AR plus a small direct AR contribution. However, as shown in Fig. \ref{cplpartI}a,
for $V_{2}\neq0$, the $ARMR$ presents negative values even for high values
of the $P_{1}$ and $P_{2}$. In fact, the potential $V_{2}$ shifts $I_{P}$
and $I_{AP}$ along the current axis, as shown in Fig. \ref{cplpartI}b. In
this case, the current $I_{P}$ can be higher than $I_{AP}$ for some range of
$V_{1}$ even if the amplitude of the former is close to zero.

The parameters controlling the amplitude and the shift of the currents with $%
V_{1}$ and $V_{2}$ are the coupling constants $\Gamma_{1}$ and $\Gamma_{2}$,
respectively. In fact, by increasing $\Gamma_{1}$ and $\Gamma_{2}$, the
admixture of  states at the ferromagnets with the dot levels is also increased.
Therefore, more electrons can be transferred to the superconductor by direct
ARs, resulting in higher amplitudes of $I_{P}$. Since the values used are $%
\Gamma_{1}=0.20$ and $\Gamma_{2}=0.80$, the amplitude of $I_{P}$ is smaller
in comparison to the shift along the current axis. On the other hand, by
comparing the figures \ref{cplpartI}b and \ref{cplpartI}c, we note that the
amplitude of $I_{AP}$ is almost independent on the relation between $%
\Gamma_{1}$ and $\Gamma_{2}$. In fact, $I_{AP}$ is carried almost through
crossed ARs which picks up one up-spin electron from $F_{1}$ and another
down-spin electron from $F_{2}$. Since the total current entering into the
superconductor must be unpolarized, it is limited by the electrode with
lower injection of electrons. The difference between these two processes
(crossed and direct AR) with respect to the variations of $\Gamma_{1}$ and $%
\Gamma_{2}$ allows the control of the sign of $ARMR$ through external
parameters.
\begin{figure*}[htb]
\centering
\includegraphics[scale=0.78]{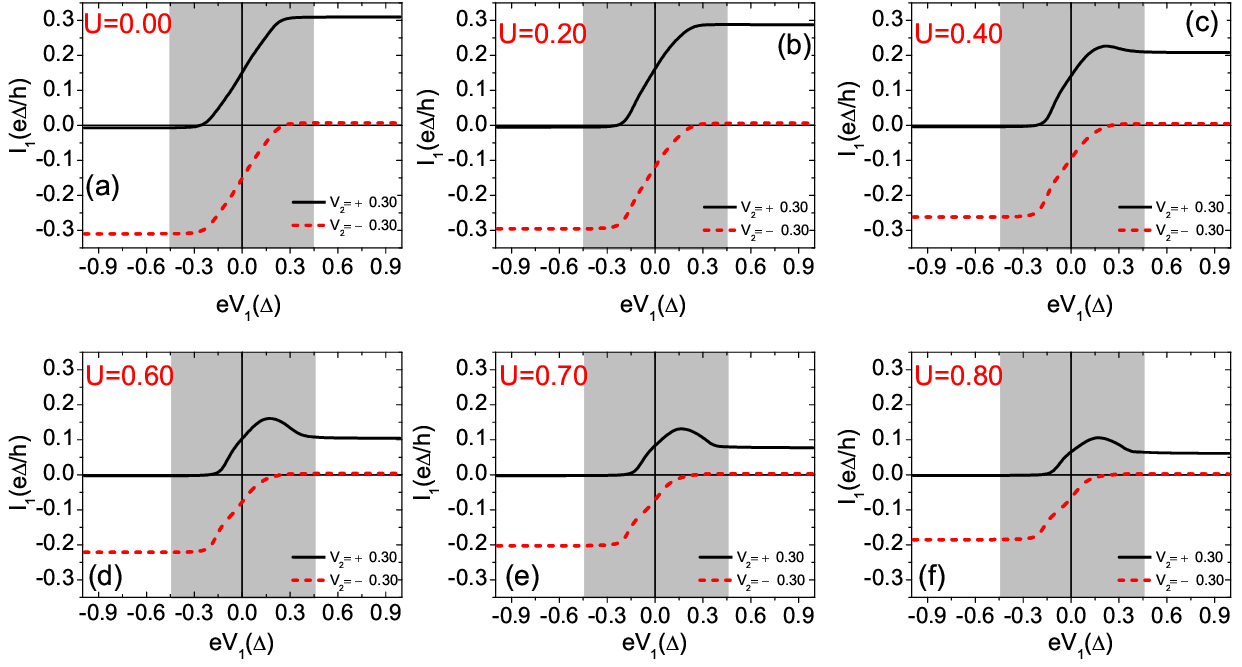}
\caption{(Color Online) Current through the terminal $F_{1}$ for $%
V_{2}=-0.30 $ (red curve) and $V_{2}=0.30$ (black curve) for different
values of the intradot interaction $\mathcal{U}$. Fixed parameters: $\protect%
\theta=\protect\pi$, $\Gamma_{1}=0.20$, $\Gamma_{2}=0.80$, $\Gamma_{s}=0.30$%
, $t_{ab}=0.20$, $V_{ga}=V_{gb}=0$, $P_{1}=P_{2}=0.95$, $k_{B}T=0.01$. All
the parameters are expressed in superconductor gap units.}
\label{switch}
\end{figure*}
\begin{figure*}[htb]
\centering
\includegraphics[scale=0.78]{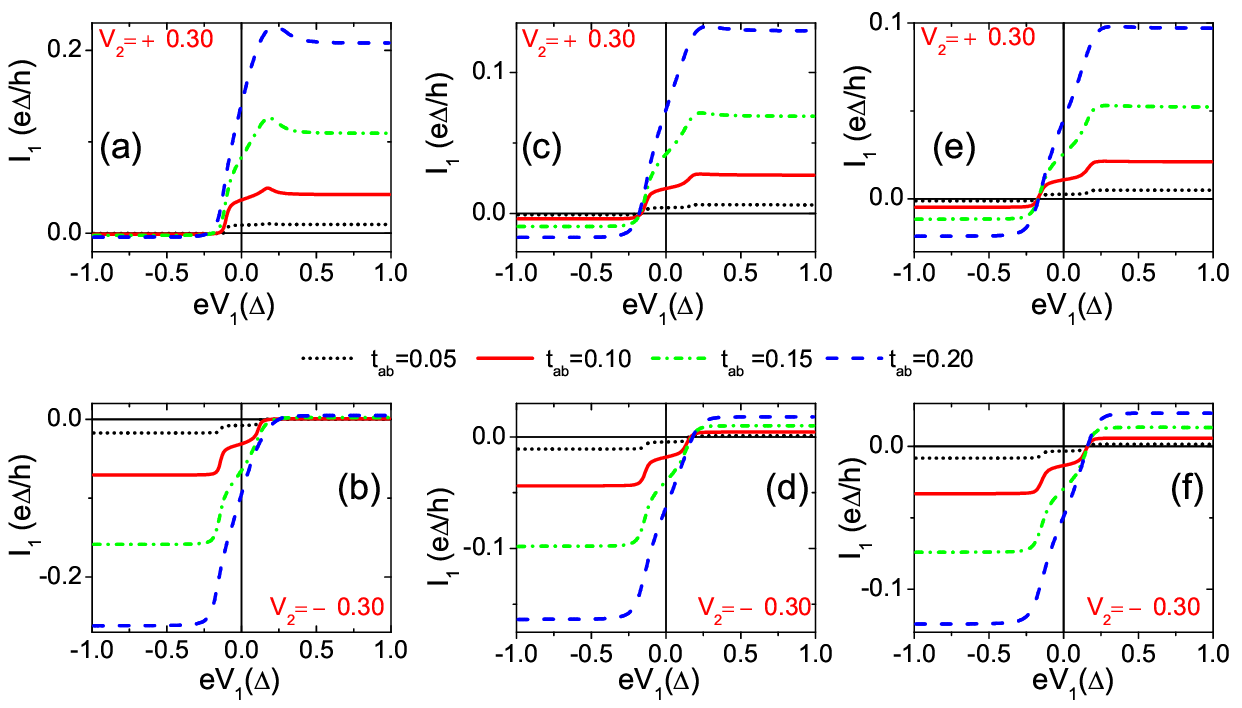}
\caption{(Color Online) Current through the terminal $F_{1}$ for different
values of hopping parameter $t_{ab}$. Figs. (a) and (b): $P_{1}=P_{2}=0.95$;
(c) and (d): $P_{1}=P_{2}=0.60$; (e) and (f): $P_{1}=P_{2}=0.24$. Fixed
parameters: $\protect\theta=\protect\pi$, $\Gamma_{1}=0.20$, $%
\Gamma_{2}=0.80 $, $\Gamma_{s}=0.30$, $\mathcal{U}=0.40$~and~$k_{B}T=0.01$.
All the parameters are expressed in superconductor gap units.}
\label{polarizationtab}
\end{figure*}
\begin{figure*}[!]
\centering
\includegraphics[scale=0.78]{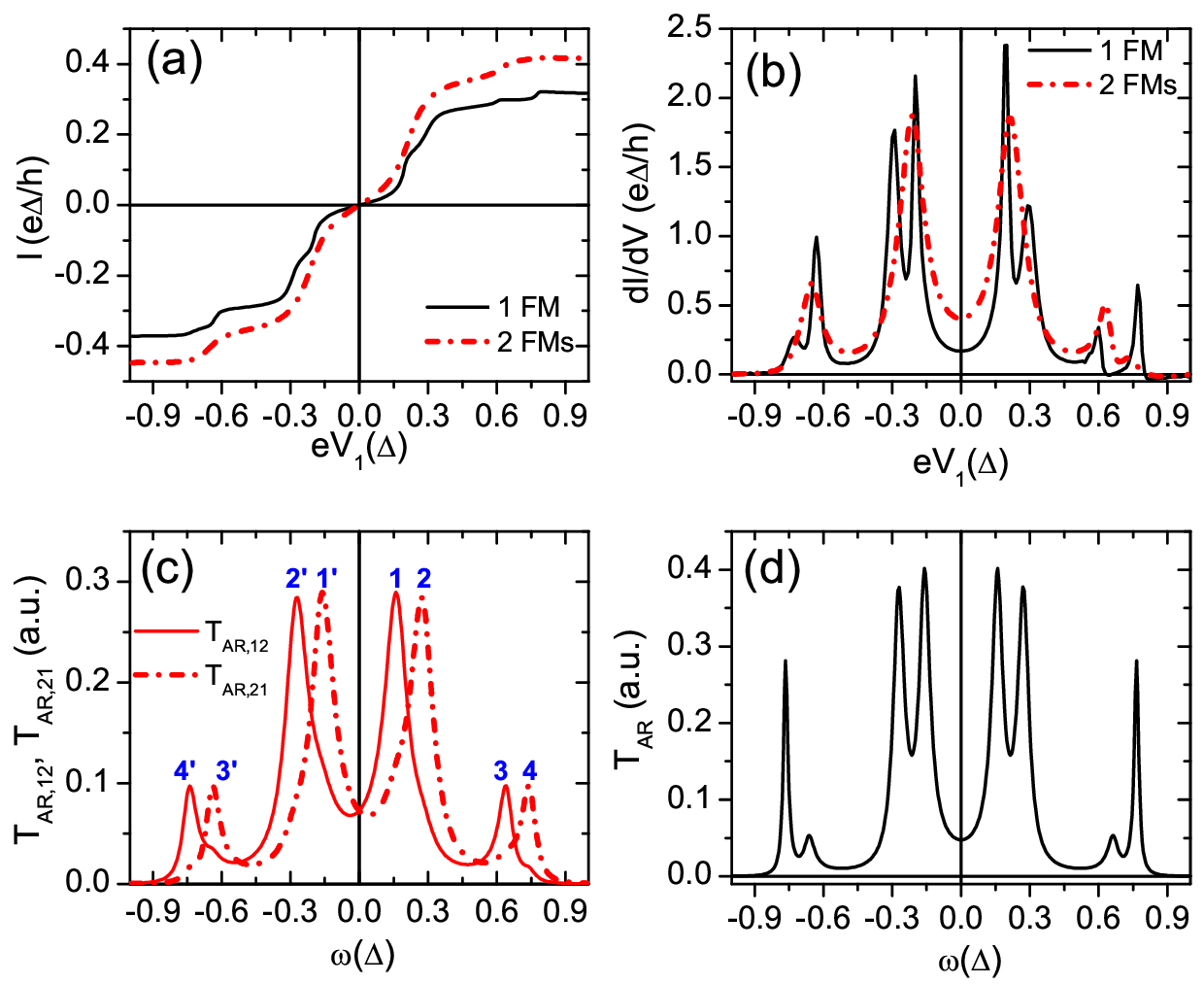}
\caption{(Color Online) (a) Currents through the one-terminal $%
F_{1}-QD_{a}-QD_{b}-S$ system (solid black curve) and the two-terminals $%
(F_{1},F_{2})-QD_{a}-QD_{b}-S$ system (red dash-dotted curve). (b)
Corresponding differential conductance curves. (c) Transmittance curves for
crossed AR of the two-terminal $(F_{1},F_{2})-QD_{a}-QD_{b}-S$ system. $%
T_{AR,12}$ represents the transmittance for an up-spin electron of $F_{1}$
to be reflected as a down-spin hole of $F_{2}$; $T_{AR,21}$ represents the
transmittance for an up-spin electron of $F_{2}$ to be reflected as a
down-spin hole of $F_{1}$. (d) Transmittance curve for the one-terminal
system $F_{1}-QD_{a}-QD_{b}-S$. Fixed parameters: $\Gamma_{1}=0.20$, $%
\Gamma_{s}=0.30$, $\mathcal{U}=0.90$, $t_{ab}=0.20$, $%
k_{B}T=0.01$, $V_{2}=0.30$. $\Gamma_{2}=0.80$ and $P_{1}=P_{2}=0.95$
for the system with
two-terminals, and $\Gamma_{2}=0$ and $P_{1}=0.50$ for the system with one ferromagnet. All
the parameters are expressed in superconductor gap units.}
\label{interference}
\end{figure*}
\begin{figure*}[!]
\centering
\includegraphics[scale=0.7]{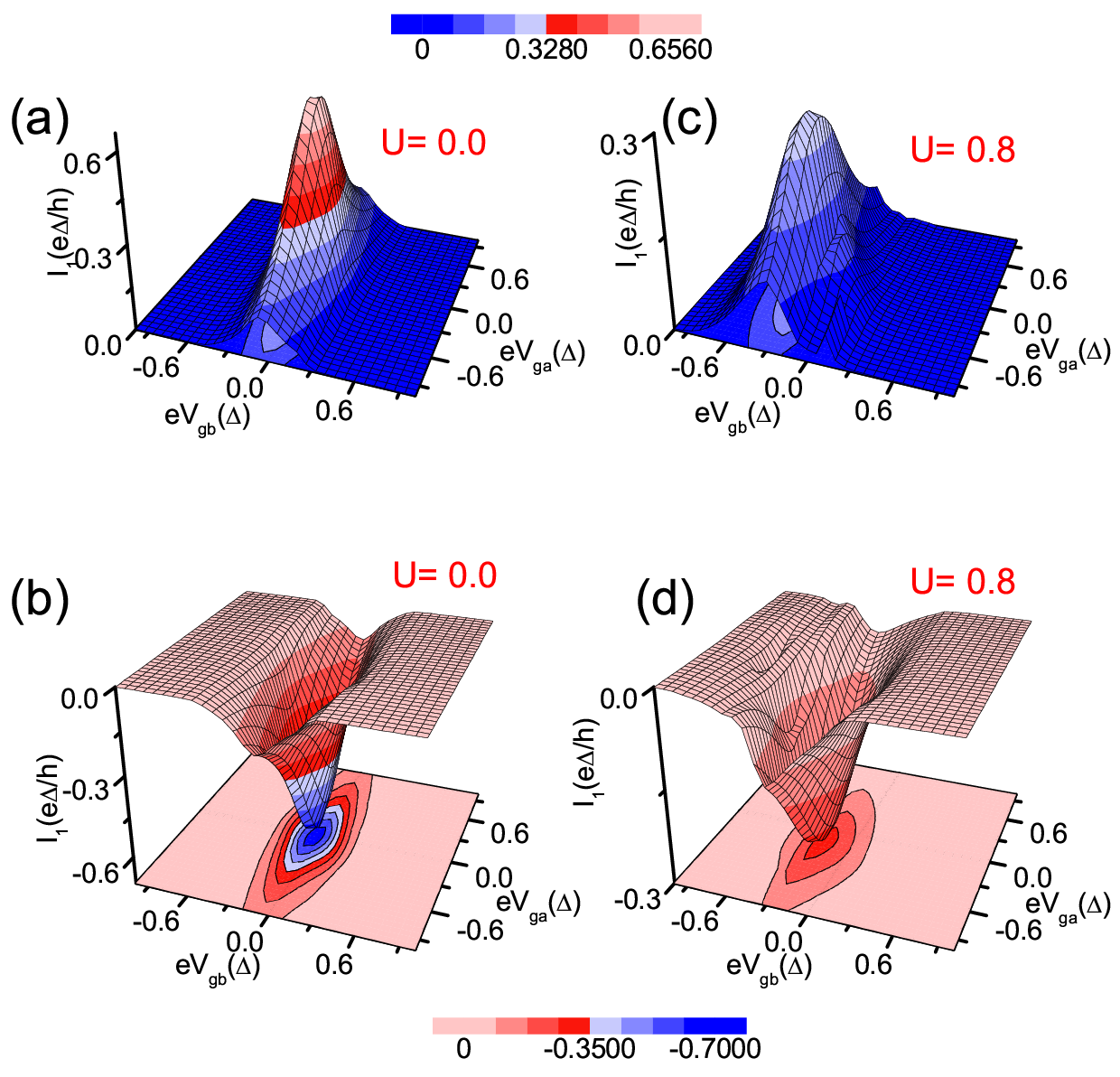}
\caption{(Color Online) Current through the terminal $F_{1}$ as a function
of the gate voltages $V_{ga}$ and $V_{gb}$. (a) $V_{1}=V_{2}=+0.30$ and $%
\mathcal{U}=0$. (b) $V_{1}=V_{2}=-0.30$ and $\mathcal{U}=0$. (c) $%
V_{1}=V_{2}=+0.30$ and $\mathcal{U}=0.8$. (d) $V_{1}=V_{2}=-0.30$ and $%
\mathcal{U}=0.80$. Fixed parameters: $\protect\theta=\protect\pi$, $%
\Gamma_{1}=0.20$, $\Gamma_{2}=0.80$, $\Gamma_{s}=0.30$, $t_{ab}=0.20$, $%
P_{1}=P_{2}=0.95$, $k_{B}T=0.01$. All the parameters are expressed in
superconductor gap units.}
\label{surfaces}
\end{figure*}

The curves for $I_{AP}$ display some interesting features. Unlike the
current $I_{P}$, the shift of $I_{AP}$ along the current axis is related to
the applied bias rather than the coupling constants $\Gamma_{1}$ and $%
\Gamma_{2}$. In fact, the zero value of the current $I_{AP}$ is found
through the condition $V_{1}=-V_{2}$ (see also Ref. \onlinecite{principal}
for one-dot case). This condition determines the shifts of the current $%
I_{AP}$ when the value of the bias is changed in the electrode $F_{2}$. The
current $I_{AP}$ is given by the sum of the currents of each electrode, $%
I_{1}$ and $I_{2}$, which present the same behavior shown by the total
current. This is a result of the coherence between the leads in the crossed
AR. To illustrate this point, in Fig. \ref{switch} the current curves in the
electrode $F_{1}$ are shown for values of the intradot interaction $\mathcal{%
U}$ ranging from 0 up to 0.80. By comparing the curves of Figs. \ref{switch}%
c and \ref{cplpartI}c, it can be noted that the amplitude of $I_{1}$ is half
the amplitude of the total current showing a balance in the contribution of
each ferromagnet. For all curves shown in Fig. \ref{switch}, the system
works as a switch when $|V_{1}|\gtrsim0.32$ (out of the shaded region): if
the bias in $F_{2}$ is changed from zero to $\pm 0.30$ the current through $%
F_{1}$ is commuted from its maximum to a value close to zero. In Fig. \ref%
{cplpartI}c, the current is reduced from 0.20 to 4.5$\times10^{-3}$ at $%
V_{1}=0.60$ as $V_{2}$ is changed from +0.30 to -0.30. This implies that the
current is reduced to 2\% of its maximum value. This small ``leakage"
current could be eliminated in the case in which the ferromagnets are fully
polarized. This switching behavior of the system can be useful in practical
applications since the system behaves as a transistor. The switching effect
persists even for high values of the intradot interaction as shown in Fig. %
\ref{switch}f, for $\mathcal{U}=0.80$. However, as the interaction
increases, an asymmetry in the curves with respect to the sign of $V_{1}$
emerges: the amplitude of $I_{1}$ is strongly reduced for $V_{1}>0$ but is
weakly reduced for $V_{1}<0$. In curves with $V_{2}=+0.30$, there is a
reduction of the current with the increase of the applied potential for $%
\mathcal{U}>0.20$. This effect has been studied by the authors in a
previous work \cite{siqueira} and its explanation is based on the appearance
of asymmetries caused by the interaction in the local density of states
(LDOS) at the QDs.

The results for the transistor based on AR depend on high values of the
polarization of the electrodes since the difference between direct and
crossed processes is the key for the behavior observed in this system. In
addition, the step-like behavior of the current stems from the
localized LDOS around the superconductor chemical potential ($\mu_{S}$).
This way, in an experimental realization of this system, a pertinent
question would be if the transistor effect persists for smaller values of
polarization and hopping between the QDs. The latter parameter being
responsible for the resonant structure of the LDOS around $%
\mu_{S}$. In order to analyze these points, in Fig. \ref{polarizationtab}
some curves for the current $I_{1}$ are presented for different values of
the ferromagnet polarizations and hopping parameter between dots. We considered three
different values of polarizations: $P_{1}=P_{2}=0.95$ (Figs. \ref%
{polarizationtab}a and \ref{polarizationtab}b); $P_{1}=P_{2}=0.60$ (Figs. %
\ref{polarizationtab}c and \ref{polarizationtab}d) and $P_{1}=P_{2}=0.24$
(Figs. \ref{polarizationtab}e and \ref{polarizationtab}f). In Figs. \ref%
{polarizationtab}a and \ref{polarizationtab}b the polarizations take the
same values as in Figs. \ref{cplpartI} and \ref{switch}, but $t_{ab}$ is ranging from 0.05
up to 0.20. As $t_{ab}$ is reduced from the value of 0.20, the
amplitude of the current is strongly reduced for both signs of $V_{2}$. In
spite of this reduction, the dependence of $I_{1}$ on $V_{1}$ is
qualitatively the same for all values of $t_{ab}$. Thus, the variation of $%
t_{ab}$ within this range of values preserves the behavior of the system as
a transistor. In Figs. \ref{polarizationtab}c and \ref{polarizationtab}d,
the polarization is reduced to $P_{1}=P_{2}=0.6$. In this case, an important
change can be observed in comparison to the curves of Figs. \ref%
{polarizationtab}a and \ref{polarizationtab}b: the leakage current displays
now a noticeable value which increases with the hopping parameter. For $%
t_{ab}=0.20$, the maximum value of the leakage current is approximately $\pm
0.02$ for $V_{1}=\pm 0.78$ and $V_{2}=\mp0.3$. As the hopping parameter is
reduced, the leakage current is also reduced as shown in Figs. \ref%
{polarizationtab}c and \ref{polarizationtab}d in which $t_{ab}$ is changed
from 0.20 to 0.05. Even though, for all curves the maximum value of the
leakage current is about $13\%$ of the current maximum for both signs of $%
V_{2}$. As the polarization is further reduced, the leakage current
increases as shown by the Figs. \ref{polarizationtab}e and \ref%
{polarizationtab}f, for $P_{1}=P_{2}=0.24$. In this case, the leakage
current is about 20\% of the maximum current. Hence, the switching effect of
the system becomes less efficient as the polarization is reduced below 80\%.
On the other hand, the reduction of $t_{ab}$ does not destroy the switching
effect since it just reduces the current amplitudes.


\begin{figure*}[!]
\centering
\includegraphics[scale=0.78]{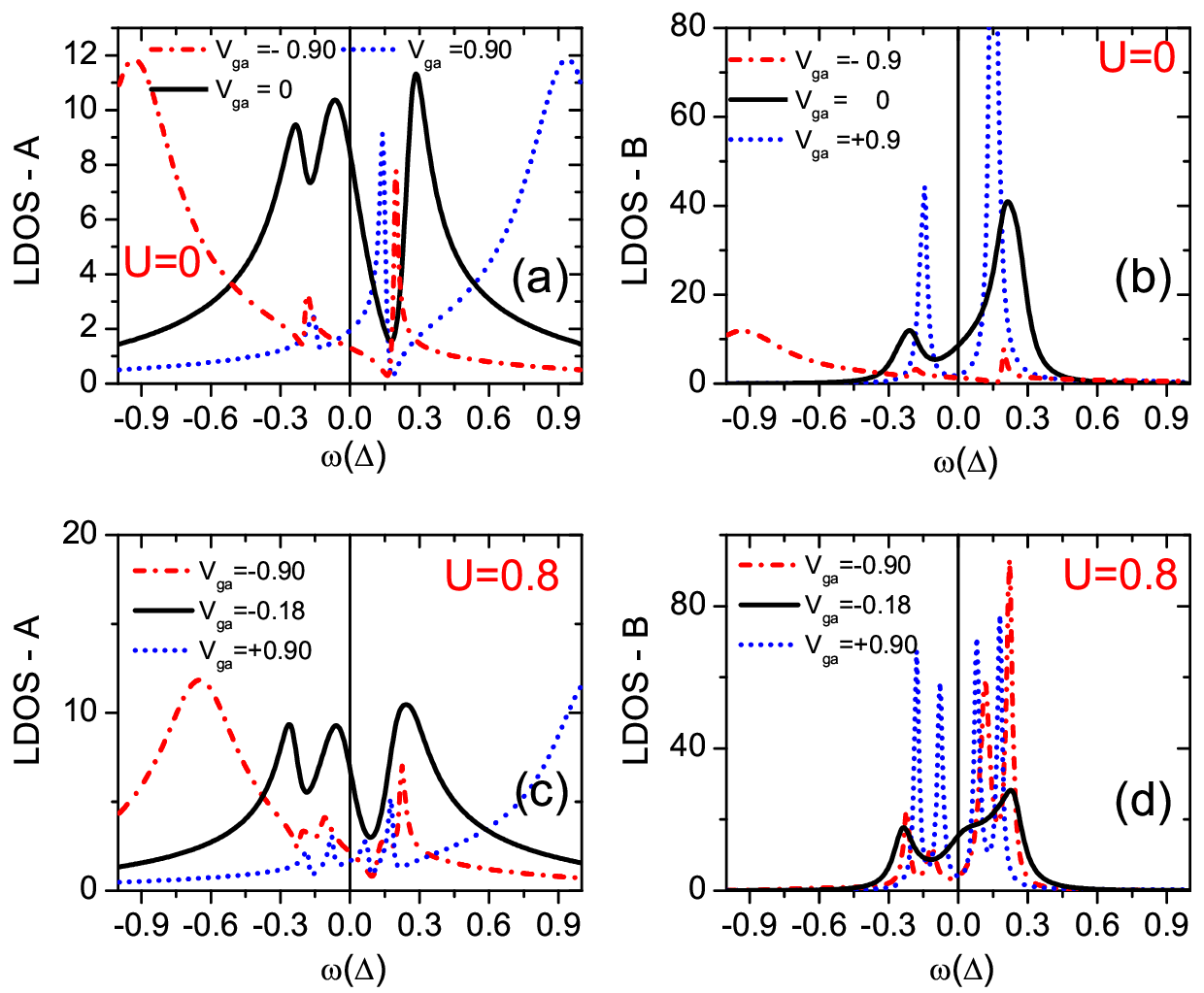}
\caption{(Color Online) Local density of states (LDOS) for the QDs for some
values of the gate voltage $V_{ga}$. LDOS-A is the density of states of the
QD connected to the ferromagnets. LDOS-B is the density of states of the
second QD, connected to the superconductor. (a) LDOS-A with $V_{gb}=0$ and $%
U=0.0$. (b) LDOS-B with $V_{gb}=0$ and $U=0.0$. (c) LDOS-A with $%
V_{gb}=-0.15 $ and $U=0.8$. (d) LDOS-B with $V_{gb}=-0.15$ and $U=0.8$.
Fixed parameters: $\protect\theta=\protect\pi$, $\Gamma_{1}=0.20$, $%
\Gamma_{2}=0.80$, $\Gamma_{s}=0.30$, $t_{ab}=0.20$, $P_{1}=P_{2}=0.95$, $%
k_{B}T=0.01$. All the parameters are expressed in superconductor gap units.}
\label{ldosfigures}
\end{figure*}

In the results shown in Figs. \ref{cplpartI} to \ref{polarizationtab}, the
intradot interaction has introduced a negative differential conductance on
the current response. However, there are other effects which can also take
place under the presence of electronic correlations at the QDs. In
particular, the intradot interaction splits the up and down-spin states
at each QD, with the corresponding splitting of peaks in the transmittance and
differential conductance. In the case in which crossed ARs are
present, the effect of~~$\mathcal{U}$ on the spin-degeneracy is more complex
in comparison to systems with two--terminals as the one studied in Ref. %
\onlinecite{siqueira}. To illustrate this point, a comparison between the
responses of the present system and the two-terminal system of Ref. %
\onlinecite{siqueira} is shown in Fig. \ref{interference}.
The solid curve
(black curve) corresponds to the total current through the system $%
F_{1}-QD_{a}-QD_{b}-S$ and the dot-dashed curve (red curve) is the total
current flowing in the $(F_{1},F_{2})-QD_{a}-QD_{b}-S$ system. For the first
system, which conducts via normal ARs, we chose a not so big value of the polarization
($P_1=0.50$), otherwise the current will be very small. For the two-terminal system,
both polarization are close to 1, and the current is mainly due to non-local crossed ARs.
The corresponding differential conductance curves are shown in Fig. \ref%
{interference}b. For the system with one ferromagnet, the current displays
eight steps corresponding to the eight peaks of the differential conductance. Note that
admixture between the QD levels with the continuum of states from the ferromagnet and the Andreev levels of the superconductor, gives rise to a four-peak structure of the LDOS at the QDs\cite{siqueira}.
Under the presence of the intradot interaction,
those four peaks are split, resulting in eight peaks in the LDOS and eight
steps in the current.
In contrast, only four peaks appear in the figure of the
differential conductance for the system with 2 ferromagnets. To
understand the difference between those responses, the key factor is
to note that the polarizations have being chosen close to unity in the two-terminal system.
As a result, the conduction is carried mainly by crossed ARs, since the
ferromagnets are in the antiparallel configuration.
The electrodes carry the
current in a coherent way, i.e., one spin-up electron from $F_{1}$ and
another spin-down electron from $F_{2}$ are combined as a Cooper pair in the
superconductor. In Fig. \ref{interference}c the transmittance curves for
crossed ARs are shown. $T_{AR, 12}$ represents the transmittance for a
spin-up electron of $F_{1}$ to be reflected as a spin-down hole in $F_{2}$,
while $T_{AR,21}$ represents the transmittance for a spin-down electron of $F_{2}$
to be reflected as a spin-up hole in $F_{1}$. It can be noted that the
curves are not symmetric with respect to the energy origin. This feature
contrasts with the transmittance in the case of one ferromagnetic electrode,
which is illustrated in Fig. \ref{interference}d. However, the two curves
displayed in (c), when combined present a symmetric
character, as one can observe by locating the peaks of each spectrum. The
peak labeled by 1 in $T_{AR, 12}$ and the peak $1^{\prime}$ in $T_{AR,21}$
are located at $\omega=+0.16$ and $-0.16$, respectively. The same symmetry
can be observed between the other pairs of peaks. This shows us that the
ferromagnets act cooperatively in the
transport, circumventing the effect of $\mathcal{U}$ in breaking spin degeneracy.
In fact, the interaction $\mathcal{U}$ is only responsible for the shifts of the
curves with respect to the origin. However, since $T_{AR,12}$ and $T_{AR,21}$
are displaced symmetrically, there is no sensible
effect on the current and differential conductance as shown in Figs. \ref%
{interference}a and \ref{interference}b. In the case with one ferromagnet,
there is only one transmittance curve, and the effects of the
interaction can be observed by the splitting of the peaks. It is important
to note that transport via AR requires electrons with energies disposed
symmetrically with respect to the superconductor chemical potential.
This is necessary in order to form Cooper pairs within the superconductor
and sustain the subgap current associated with AR.
Next, we consider the effect of the gate potentials on the electric
transport of the system. In Fig. \ref{surfaces}, the current through $F_{1}$
($I_{1}$) is plotted in terms of the gate potentials $V_{ga}$ and $V_{gb}$
applied on the QDs $a$ and $b$, respectively. The bias in the ferromagnets
are fixed at $V_{1}=V_{2}=\pm 0.30$ corresponding to the maximum value of $%
I_{1}$ for the curves of Fig. \ref{polarizationtab}a and \ref%
{polarizationtab}b. In Figs. \ref{surfaces}a and \ref{surfaces}b the
interaction at the QDs is zero and $I_{1}$ displays a single peak centered
at $V_{ga}=V_{gb}=0$. The same behavior is observed in Fig. \ref{surfaces}b
for $V_{1}=V_{2}=-0.30$. When the interaction at the QDs is present, the
current still exhibits a single peak as observed in Figs. \ref{surfaces}c
and \ref{surfaces}d for $\mathcal{U}=0.80$. However, the peak is now located
at $V_{ga}=V_{gb}=-0.18$ and its amplitude has been reduced to half of the
value for $\mathcal{U}=0.0$. Therefore, under the presence of interaction at
the QDs, gate voltages must be used in order to find the maximum condition
for the electrical current. By the projections on the plane $V_{gb}\times
V_{ga}$ in Figs. \ref{surfaces}b and \ref{surfaces}d, it can be noted that
the variation of $I_{1}$ is asymmetric with respect to the gate potentials.
In fact, $I_{1}$ is different from zero in the entire range of $V_{ga}$ but
is appreciable only within a very narrow range of $V_{gb}$.
This can be explained by noting that the system is not symmetric, with each QD
connected to a different electrode, and subjected to a different hybridization of
quantum states. This is also reflected in the structure of the LDOS in both QDs,
as shown in Fig.\ref{ldosfigures}.
By changing the gate voltages $V_{ga}$ and $V_{gb}$, it is possible to change
the QDs levels and the LDOS. In simpler systems, in which the coupling
between the QDs and the electrodes is weak, the gate voltage just shifts the QDs
levels with respect to the chemical potential of the electrodes. Hence, the
behavior of the LDOS can be described in a intuitive manner, being possible
to relate the changes of the current directly with those of the levels of
the QDs. In our example, the analysis is subtler, since the QD
levels are admixed with the continuous band of the ferromagnets and the discrete
Andreev levels of the superconductor. This way, the effects of the
gate voltages  on the LDOS are more complex to resolve.
To illustrate this point, in Fig. \ref{ldosfigures}
we display some curves for the LDOS of both QDs, for different
values of $V_{ga}$ and $V_{gb}$. The LDOS are obtained by standard methods,
from the imaginary part of elements $11$ (dot~$a$) and $33$ (dot~$b$) of the retarded
Green function of these QDs. Explicit formulae can be found in Ref. \onlinecite{siqueira}.

The curves correspond to the bias $V_{1}$ and $V_{2}$ equal to $0.30$
meaning that only the states within the range of conduction $%
-0.30<\omega<+0.30$ are contributing to the transport. As the gate
voltages are varied, the LDOS within this window changes, modifying the
response of the system to applied voltages in $F_{1}$ and $F_{2}$. For the
non-interacting case, the maximum value of the current occurs for $%
V_{ga}=V_{gb}=0$. The corresponding LDOS for both QDs are represented by the
solid curves (black curves) in Figs. \ref{ldosfigures}a and \ref{ldosfigures}%
b. It can be noted that the LDOS is spread over the entire range $%
-0.30<\omega<+0.30$ which implies that the current is carried almost over
all range determined by the applied bias. For $V_{ga}=-0.90$ (blue dotted
curve) and $V_{ga}=+0.90$ (red dash-dotted curve) LDOS-A exhibits a maximum
at $\omega=-0.93$ and $+0.93$, respectively. In the range of interest, there
are two well localized peaks but with asymmetric amplitudes. These two peaks
also appear in the curves for LDOS-B as shown in Fig. \ref{ldosfigures}b.
However, these peaks present higher amplitudes for $V_{ga}=+0.90$, being very
suppressed for $V_{ga}=-0.90$. This asymmetric pattern explains the
suppression of the current for these values of the gate potentials. As
pointed in our previous work \cite{siqueira}, the symmetry of the peaks in the
LDOS is crucial for the transport, since states located at opposite
values of the energy combine to form Cooper pairs in the superconductor.
If one of these peaks is suppressed, the
effective number of states participating in the conduction process is
effectively reduced and the current becomes smaller. For the interacting
case, shown in Figs. \ref{ldosfigures}c and \ref{ldosfigures}d, the intradot
interaction splits the peaks of the LDOS at both QDs. However, since some
peaks are strongly suppressed, we do not see them within the scale of the graph.
As an example, take the value $V_{ga}=\pm0.90$. For the QD
coupled to the superconductor (dot $b$), the LDOS exhibits four peaks localized
inside the range $-0.30<\omega<+0.30$. However, the corresponding states for
the QD coupled to the ferromagnets (dot $a$) are completely suppressed in the same
range (LDOS-A displays only four small peaks around $%
\omega=0$). In contrast, when $V_{ga}=-0.18$, both LDOS are
appreciable inside the range of conduction, which explains the maximum value
of the current shown in Figs. \ref{surfaces}c and \ref{surfaces}d.

The system is very sensible to variations of the gate potentials, as shown
by the results in Figs. \ref{surfaces} and \ref{ldosfigures}. In fact, the
voltage values involved are restricted to the superconductor gap, and small
variations of the parameters within this range are sufficient to change the
transport response of the system. See, for instance, the cases of Figs.
\ref{surfaces}c and \ref{surfaces}d, for fixed $V_{gb}=-0.18$. Changing $V_{ga}$
from -0.18 to 0, reduces the current to 38\% of its maximum value.

\section{Conclusion}
The combination of
superconductivity with ferromagnetism in nanostructures gives rise to
most interesting properties, probably useful in future technologies.
In this work we have studied the magnetoresistance and the current
properties of the $(F_{1},F_{2})-QD_{a}-QD_{b}-S$ hybrid system, in the case
of subgap currents, when the transport is solely due to AR processes.
We found that the magnetoresistance sign can be switched by applying an external potential in one of the ferromagnetic leads. In addition, the current carried by crossed ARs can also be controlled through the potential of the ferromagnets. Being a nonlocal process, crossed ARs allow control of the current in one ferromagnet, say $F_{1}$, by means of the potential applied to the other, say $F_2$, with the system behaving as a switch for some values of the parameters.
The switching effect works better for polarization values close to unity. In
fact, the leakage current in the inverse direction is completely suppressed
when the ferromagnets are fully polarized. High polarizations ($>$%
90\%) values have been observed in ferromagnetic films of $CrO_{2}$ by
Soulen Jr. and co-workers~\cite{polarization}. Some high values ($>$85\%)
have  also been reported in ferromagnetic semiconductors based on $GaMnAs$ \cite%
{polarization2}. This way, representative values used in our numerical calculations
could be implemented in experiments. Inclusion of the intradot interaction $\mathcal{U}$ does not kill the switching effect, as shown by examples in Fig. \ref{switch}.
However, it is worth mentioning that our results were obtained from a mean field approximation in treating correlations at the QDs, with fluctuations being neglected.
Important effects, such as the negative differential conductance and the lifting of spin degeneracy, could be washed out by fluctuations, and we have to look for a safe ground in order to apply mean field results. Qualitatively, this domain corresponds to high polarization values and nonzero gate voltages, which strongly suppress fluctuations. However, the exact extension of the validity of the approximation used in this work can be addressed only by experiments.

The switching property shown by the $(F_{1},F_{2})-QD_{a}-QD_{b}-S$ system
resembles the conventional transistors used in large scale in any commercial
electronic device. The future of the electronics in the nanometer domain demands
devices which mimic the conventional ones, and the system presented in this
work may be a contribution in this direction.

\bibliography{referencias}

\begin{thebibliography}{64}
\bibitem [{noa()}]{noanomalous}%
  \BibitemOpen
  \href@noop {} {\ }\bibinfo {note} {As discussed previously, we are assuming
  conditions that exclude `anomalous' AR, that is reflections with spin flip.
  In this case, the reflected hole flips its spin and enters the same
  ferromagnetic lead of the incident electron, inducing a triplet proximity
  effect. This local reflection will compete with the nonlocal crossed AR.
  However, spin-flip scattering is only possible in the presence of
  inhomogeities at the ferromagnets.}\BibitemShut {Stop}%
\end{thebibliography}

\end{document}